\documentclass[superscriptaddress,twocolumn]{revtex4}
\usepackage{amssymb}
\usepackage{natbib}
\usepackage{amsmath}
\usepackage{amsfonts}
\usepackage{graphicx}
\usepackage{mathrsfs}
\usepackage{dcolumn} 
\usepackage{bm}
\usepackage{color}
\usepackage{cancel}
\usepackage{mathbbol}
\usepackage{epsfig}
\usepackage{units}
\usepackage{esint}
\usepackage{soul} 
\usepackage{url}

\definecolor{rot}{rgb}{0.75,0.05,0.25}
\definecolor{hellgrau}{gray}{0.5}
\definecolor{blau}{rgb}{0,0,0.7}

\def\Tr{\mbox{Tr}}

\newtheorem{theo}{Theorem}
\newtheorem{state}{Statement}

\begin{document}

\title{Quantum Measurement Cooling}
\author{Lorenzo Buffoni}
\affiliation{\mbox{Department of Information Engineering, University of Florence,} via S. Marta 3, I-50139 Florence, Italy}
\affiliation{\mbox{Department of Physics and Astronomy, University of Florence,} via G. Sansone 1, I-50019 Sesto Fiorentino (FI), Italy}
\author{Andrea Solfanelli}
\affiliation{\mbox{Department of Physics and Astronomy, University of Florence,} via G. Sansone 1, I-50019 Sesto Fiorentino (FI), Italy}
\author{Paola Verrucchi}
\affiliation{\mbox{Istituto dei Sistemi Complessi, Consiglio Nazionale delle Ricerche,
via Madonna del Piano 10, I-50019 Sesto Fiorentino (FI), Italy}}
\affiliation{\mbox{Department of Physics and Astronomy, University of Florence,} via G. Sansone 1, I-50019 Sesto Fiorentino (FI), Italy}
\affiliation{\mbox{INFN Sezione di Firenze, via G.Sansone 1, I-50019 Sesto Fiorentino (FI), Italy}}
\author{Alessandro Cuccoli}
\affiliation{\mbox{Department of Physics and Astronomy, University of Florence,} via G. Sansone 1, I-50019 Sesto Fiorentino (FI), Italy}
\affiliation{\mbox{INFN Sezione di Firenze, via G.Sansone 1, I-50019 Sesto Fiorentino (FI), Italy}}
\author{Michele Campisi}
\affiliation{\mbox{Department of Physics and Astronomy, University of Florence,} via G. Sansone 1, I-50019 Sesto Fiorentino (FI), Italy}
\affiliation{\mbox{INFN Sezione di Firenze, via G.Sansone 1, I-50019 Sesto Fiorentino (FI), Italy}}
\affiliation{\mbox{Kavli Institute for Theoretical Physics, University of California, Santa Barbara, CA 93106, USA}}

\begin{abstract}
Invasiveness of quantum measurements is a genuinely quantum mechanical feature that is not necessarily detrimental: Here we show how quantum measurements can be used to fuel a cooling engine. We illustrate quantum measurement cooling (QMC) by means of a prototypical two-stroke two-qubit engine which interacts with a measurement apparatus and two heat reservoirs at different temperatures. We show that feedback control is not necessary for operation while entanglement must be present in the measurement projectors. We quantify the probability that QMC occurs when the measurement basis is chosen randomly, and find that it can be very large as compared to the probability of extracting energy (heat engine operation), while remaining always smaller than the most useless operation, namely dumping heat in both baths. These results show that QMC can be very robust to experimental noise. A possible low-temperature solid-state implementation that integrates circuit QED technology with circuit quantum thermodynamics technology is presented.
\end{abstract}

\maketitle

\begin{figure}[b]
\includegraphics[width=\linewidth]{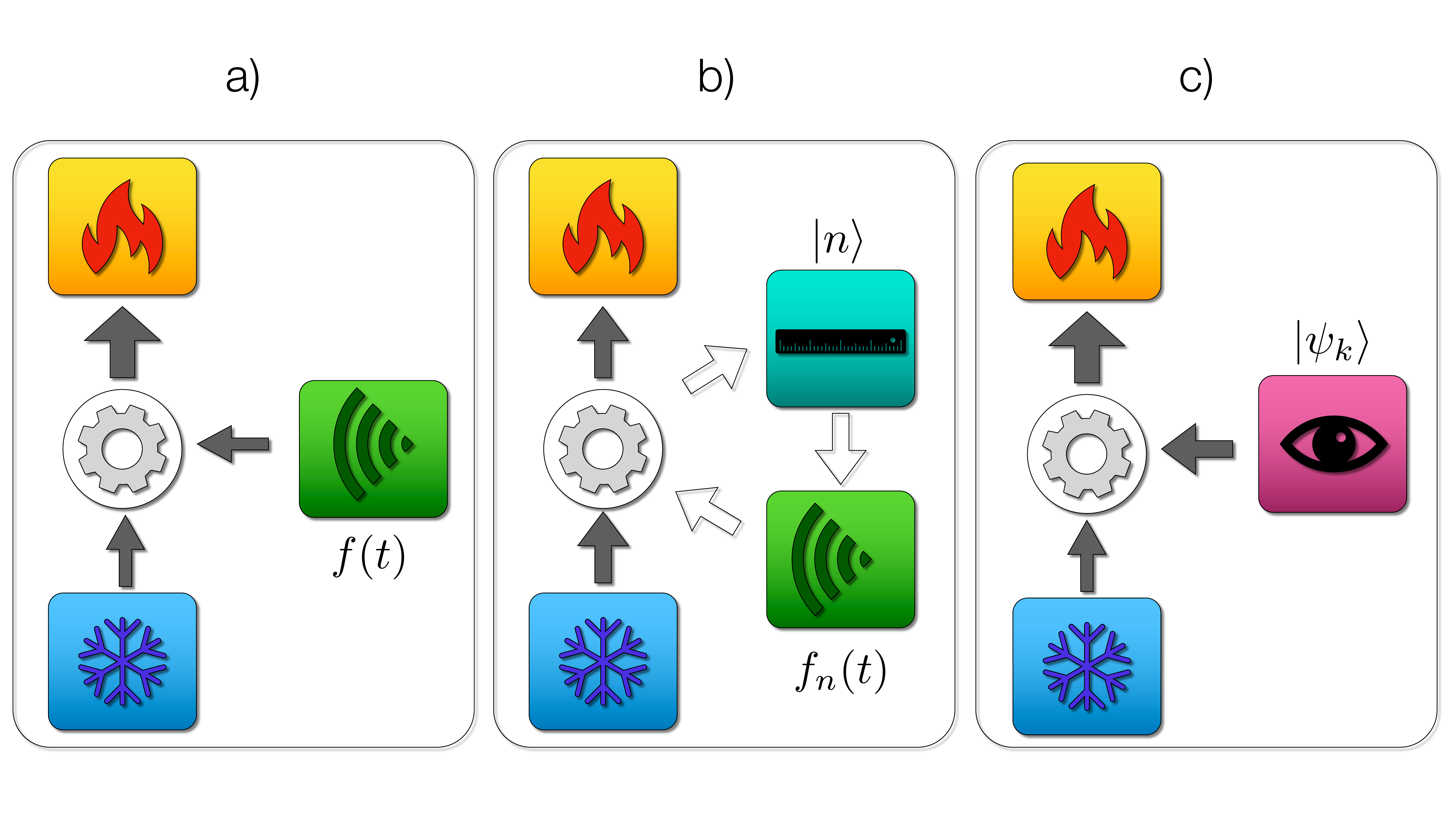}
\caption{Various ways to pump a heat current from a cold to a hot reservoir. (a) In standard refrigeration the heat current is powered by energy injected by a time dependent driving force $f(t)$. (b) In Maxwell demon refrigeration heat current is generated by a feedback loop where various driving forces $f_n(t)$ are applied depending on the outcome, $n$, of non-invasive measurements on the working substance, without energy injection. (c) In quantum measurement cooling, put forward here, the heat current is powered by energy provided via invasive measurements on an appropriate measurement basis $\{|\psi_k\rangle\}$, without performing feedback control. Solid arrows represent flow of energy.}
\label{fig:1}
\end{figure}

\emph{Introduction.--} The second law of thermodynamics dictates that heat naturally flows from hot bodies to cold ones \cite{Fermi56Book}. There are two standard ways to intervene and reverse the natural flow of heat (see Fig. \ref{fig:1}): (a) use work supplied by an external time-dependent driving force $f(t)$ thus realizing a standard refrigeration machine, see e.g., Refs. \cite{Campisi15NJP17,Campisi16JPA49a};  (b) implement a     Maxwell demon that steers the heat by means of a feedback control loop, consisting in acquisition of information about the state $|n\rangle $ of the working substance by means of non-invasive measurement, followed by the timely application of various driving forces $f_n(t)$, depending on the measurement outcome, that do not do work on the system \cite{Maruyama09RMP81,Campisi17NJP19,Koski14PRL113}. 
By non-invasive measurement here we mean that the measurement basis coincides with the basis in which the state of the measured system is diagonal (in the present work that is the energy eigenbasis).
Here we will demonstrate yet another mechanism that is genuinely quantum mechanical, namely (c) to use invasive quantum measurements as a resource, in fact a fuel, that powers refrigeration, without any feedback control. We shall call this mechanism ``quantum measurement cooling'' QMC. 
QMC is performed by a demon who needs not be intelligent. It rather needs to be knowledgeable, that is it has to know which measurement basis $\{|\psi_k\rangle \}$ to employ in order that QMC occurs.

While the idea of using measurement apparata to fuel engines  is currently emerging as a new paradigm in quantum thermodynamics \cite{Ding18PRE98,Biele17QUANTMAT2,Elouard17PRL118,Elouard18PRL120}, attention has never been posed before on whether it can be used for cooling, nor on the fact that, as we elucidate below, feedback control is not necessary for exploiting the quantum-measurement fuel. We address these questions by means of a thorough investigation of a prototypical two-qubit engine \cite{Campisi15NJP17,Campisi16JPA49a,Campisi17NJP19}. Our results shed new light on many facets of the second law of thermodynamics. For example, it emerges that in order for the device to work the measurement basis must contain entangled projectors, while maximal efficiency is achieved when the post measurement statistical mixture $\rho'$  (see Eq. (\ref{eq:rho'}) below) is un-entangled. We also find that, when the measurement basis   is chosen randomly, 
the least useful operation -- i.e., dumping heat in both baths-- is the most likely outcome (hence easier to realize in practice), which conforms to intuition. Also, while energy extraction is typically very unlikely, refrigeration can be very likely.
This says that our demon needs not be very knowledgeable in order to realize QMC, or in more concrete terms, QMC can be very robust to experimental noise, that is, it is practically feasible. In the following we shall comment on a possible experimental realization.

\begin{figure}[t]
\includegraphics[width=\linewidth]{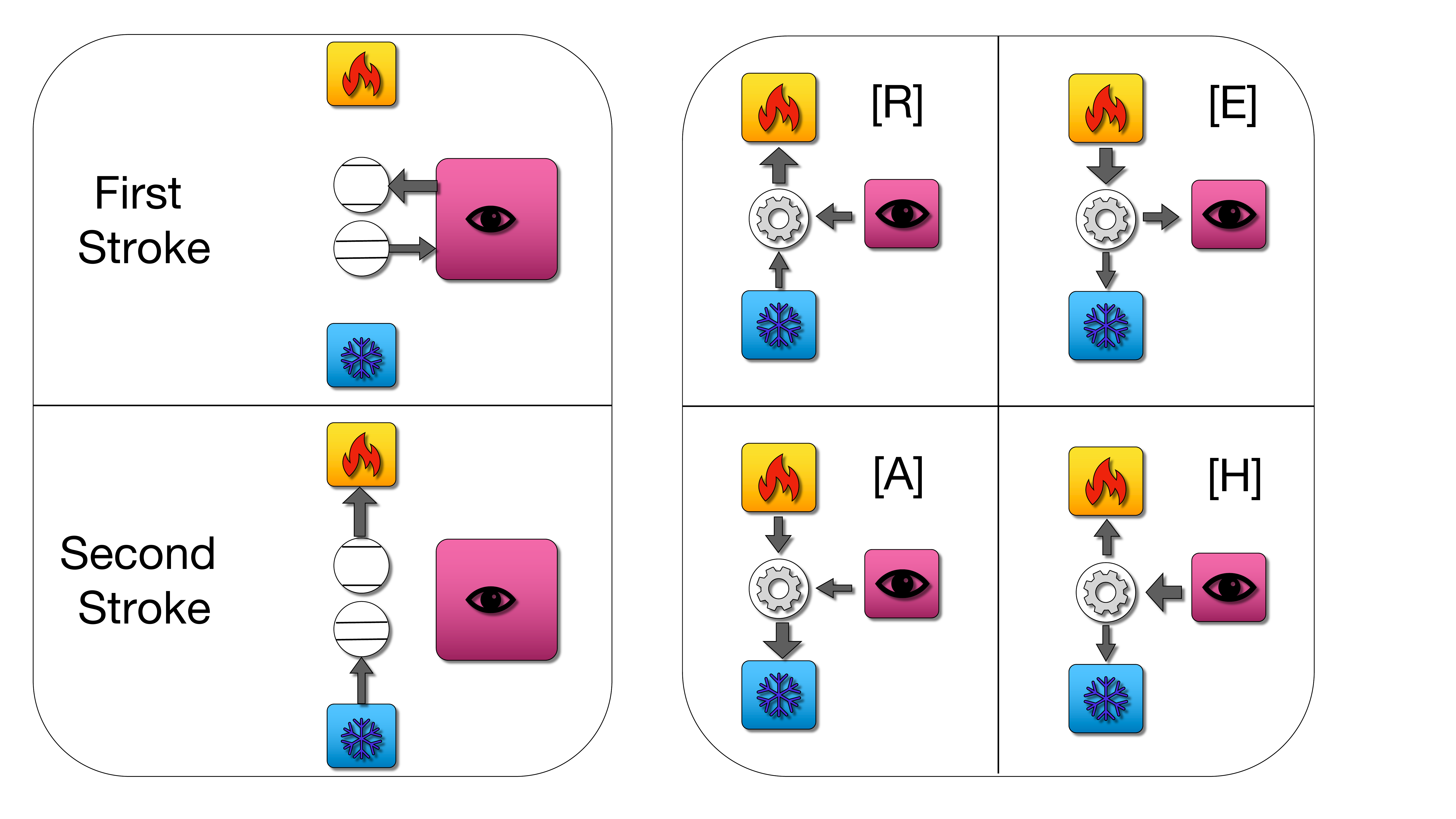}
\caption{Left panel: Two-stroke two-qubit quantum measurement cooling. During the first stroke (top) the two qubits interact with the measurement apparatus, as a consequence qubit 1 receives energy ($\langle \Delta E_1 \rangle \geq 0$), while qubit 2 loses energy ($\langle \Delta E_2 \rangle \leq 0$) with an overall positive energy injection ($\langle \Delta E_1 \rangle + \langle \Delta E_2 \rangle= \langle \Delta E \rangle \geq 0$). During the second stroke qubit 1 releases energy to the hot bath while qubit 2 withdraws energy from the cold bath. Right panel: the four possible operations allowed by the second law of thermodynamics, Eq. (\ref{eq:2ndlaw}), and energy conservation.}
\label{fig:2}
\end{figure}

\emph{The model.--}
Our model is a two-qubit engine \cite{Lloyd97PRA56,Quan07PRE76,Allahverdyan08PRE77,Campisi15NJP17,Campisi16JPA49a,Campisi17NJP19} see Fig. (\ref{fig:2}). 
Let
\begin{align}
H_i = \frac{\hbar \omega_i}{2} \sigma_z^i
\end{align}
denote the Hamiltonian of qubit $i$ expressed in terms of its Pauli matrix $\sigma_z^i$ and its resonance frequency $\omega_i$. Let
\begin{align}
H=H_1 +H_2= \sum_n E_n \Pi_n
\end{align}
be the total Hamiltonian, $E_n$ its eigenvalues with corresponding eigenprojectors $\Pi_n=| n\rangle \langle n|$ and eigenvectors $|n\rangle$. 

The two qubits are prepared each by thermal contact with a thermal bath at positive inverse temperatures $\beta_1$ and $\beta_2$ respectively, so that the initial state reads
\begin{align}
\rho = \frac{e^{-\beta_1 H_1}}{Z_1 }\otimes\frac{e^{-\beta_2 H_2}}{Z_2 } \, ,
\end{align}
where $Z_i=\Tr\, e^{-\beta_i H_i}$ is the canonical partition function.
Without loss of generality we shall set $0<\beta_1 <\beta_2$ in what follows (bath 1 hotter than bath 2).

The quantum measurement cooling cycle is illustrated in Fig. (\ref{fig:2}). In the first stroke the two-qubit system interacts with a measurement apparatus, whose effect is to erase all coherences of the two qubit compound state in the measurement basis $\{ |\psi_k\rangle \}$.
In the following we shall focus for simplicity on the case of projective measurements onto 1-dimensional sub-spaces, pointing out on a case by case basis those results that have broader validity. Denoting the projectors onto the measurement basis as $\pi_k= |\psi_k\rangle \langle  \psi_k|$ the post-measurement state 
$\rho'$ reads
\begin{align}
\rho' = \Phi[\rho] = \sum_k \pi_k \rho \pi_k \, .
\label{eq:rho'}
\end{align}
Let $\langle \Delta E_i \rangle= \Tr H_i (\Phi[\rho]-\rho)$ denote the change in the expectation value of energy of qubit $i$. Due to the property of $\Phi$ of being a unital map [namely $\Phi[\mathbb{1}]=\mathbb{1}$], it follows that \cite{Campisi17NJP19}
\begin{align}
\beta_1 \langle \Delta E_1 \rangle + \beta_2 \langle \Delta E_2 \rangle \geq 0\, ,
\label{eq:2ndlaw}
\end{align}
which expresses the second law of thermodynamics.

In the second stroke each qubit is put back in contact with its thermal bath, which restores it to its initial Gibbs state and closes the cycle. Note that in the thermalization stroke, on average, each qubit releases the energy $ \langle \Delta E_i \rangle$, gained during the first stroke, to its respective bath. The $ \langle \Delta E_i \rangle$'s represent therefore the heat exchanged with the two baths.

The sum $\langle \Delta E \rangle=\langle \Delta E_1 \rangle+ \langle \Delta E_2 \rangle$ (sometimes referred to as ``quantum heat'' \cite{Elouard17NPJQI3}) representing the energy given by the measurement apparatus is generally different from zero. Looking at the signs of the three energy exchanges $\langle \Delta E \rangle$, $\langle \Delta E_1 \rangle$, $\langle \Delta E_2 \rangle$, out of the 8 possible combinations only 4 are allowed by Eq. (\ref{eq:2ndlaw}), the condition $\langle \Delta E \rangle=\langle \Delta E_1 \rangle+ \langle \Delta E_2 \rangle$, and the condition $0<\beta_1<\beta_2$:

\begin{align}
\begin{array}{llll}
$[R]$: & \langle \Delta E_1 \rangle \geq 0 & \langle \Delta E_2 \rangle \leq  0 & \langle \Delta E \rangle \geq 0 \\
$[E]$: & \langle \Delta E_1 \rangle \leq 0  & \langle \Delta E_2 \rangle \geq 0 & \langle \Delta E \rangle \leq  0 \\
$[A]$: & \langle \Delta E_1 \rangle \leq 0  & \langle \Delta E_2 \rangle \geq 0 & \langle \Delta E \rangle \geq 0 \\
$[H]$: & \langle \Delta E_1 \rangle \geq 0  & \langle \Delta E_2 \rangle \geq 0 & \langle \Delta E \rangle \geq 0 \, .
\end{array}
\end{align}

They correspond to (see Fig. \ref{fig:2}) [R] Refrigerator: heat flows from the cold bath to hot bath, with energy injection from the measurement apparatus; [E] energy Extraction (heat engine): part of the energy naturally flowing from the hot bath to the cold bath is derouted towards the measurement apparatus); [A] thermal Accelerator: the measurement apparatus provides energy to facilitate the natural flow from the hot bath to the cold bath; [H] Heater: both baths receive energy from the measurement apparatus. Which of the 4 possibilities is realized depends on the measurement basis $\{|\psi_k\rangle \}$. The above argument holds as well for higher rank projectors.

\emph{Results.--} Our first main result is that depending on the problem parameters, some among the four possibilities, [R], [E], [A], [H], are excluded. In particular, for
$0\leq \omega_2/\omega_1 \leq \beta_1/\beta_2$ only [R] and [H] are allowed. For $\beta_1/\beta_2\leq  \omega_2/\omega_1 \leq 1$ only [E], [A], and [H] are allowed. For $\omega_2/\omega_1 \geq 1$ only [A] and [H] are allowed. Note that the most useless operation, [H], may occur in the full parameter range.  For simplicity we shall call the range $0\leq \omega_2/\omega_1 \leq \beta_1/\beta_2$ the [R]-range, the range $\beta_1/\beta_2\leq  \omega_2/\omega_1 \leq 1$ the [E]-range, and the range $\omega_2/\omega_1 \geq 1$ the $[A]$ range. In the Supplemental Material \cite{suppl} we provide a proof and discuss how this result is related to the concept of ergotropy \cite{Allahverdyan04EPL67}.

Our second main result is that, in the [E]-range, among all possible choices of measurement basis $\{|\psi_k\rangle \}$, the singlet-triplet basis
\begin{align}
\label{eq:opt-basis}
\left\{
\begin{array}{ll}
|\psi_1^*\rangle = |\uparrow \uparrow \rangle; &
|\psi_2^*\rangle =  \frac{ |\uparrow \downarrow\rangle + |\downarrow \uparrow\rangle}{\sqrt{2}} \\
|\psi_3^*\rangle = \frac{ |\uparrow \downarrow\rangle  - |\downarrow \uparrow\rangle}{\sqrt{2}};  &
|\psi_4^*\rangle =  |\downarrow \downarrow\rangle
 \end{array}
 \right.
\end{align}
maximizes the energy extraction.
This choice maximizes as well the heat engine efficiency $\eta^{[E]}= \langle \Delta E \rangle/\langle \Delta E_1 \rangle$. Similarly, in the [R]-range, the same choice of basis maximizes the energy withdrawn from the cold bath $-\langle \Delta E_2 \rangle$ and the refrigeration efficiency $\eta^{[R]} = -\langle \Delta E_2 \rangle/\langle \Delta E \rangle $. These results also show that the set of measurement-bases realizing the [E] and [R] operations are not empty, that is energy extraction and quantum measurement cooling are possible. The proof is presented in \cite{suppl}.
 
As shown in \cite{suppl} when a two-qudit working substance is considered the generic form of the optimal basis is such that it contains only factorized states of the type $|a,a\rangle$ and pairs of entangled states of the type $(|a,b\rangle \pm |c,d\rangle)/\sqrt{2}$.

When the measurement basis is $\{|\psi_k^*\rangle\}$, the expression for the $\langle \Delta E_i \rangle$'s is
\begin{align}
\langle \Delta E_{1,2} \rangle= \frac{\pm \omega_{1,2}}{2} \left ( \frac{1}{1+e^{\beta_1\omega_1}}- \frac{1}{1+e^{\beta_2\omega_2}}  \right)\, ,
\label{eq:deltaE's}
\end{align}
that is half the value obtained when implementing standard refrigeration on the two-qubit engine by means of a full $\text{SWAP}$ driving gate \cite{Campisi15NJP17}, which maximizes standard refrigeration (or energy extraction, depending on the range) over all possible unitary gates \cite{Allahverdyan08PRE77},\cite{suppl}.
We note that the same energy exchanges in Eq. (\ref{eq:deltaE's}), hence maximal efficiency, can be obtained as well with higher rank projectors, e.g., with $q_1= |\psi_1^*\rangle \langle \psi_1^*|+ |\psi_2^*\rangle \langle \psi_2^*|$, $q_2= |\psi_3^*\rangle \langle \psi_3^*|+ |\psi_4^*\rangle \langle \psi_4^*|$, or with $q_1= |\psi_2^*\rangle \langle \psi_2^*|$, $q_2= |\psi_1^*\rangle \langle \psi_1^*|+|\psi_3^*\rangle \langle \psi_3^*|+ |\psi_4^*\rangle \langle \psi_4^*|$.

In the general case of a working substance composed of two-qudits, in order for any operation other than [H] to occur some of the measurement projectors must be entangled, regardless of their rank \footnote{To see it  consider the post measurement state $\rho' = \sum_k \pi_k \rho \pi_k $ and note that it is a convex combination of the projectors $\pi_k$, $\rho' = \sum_k \pi_k r_k$. If the projectors $\pi_k$ are factorized $\pi_k = \pi^1_{k_1} \otimes \pi^2_{k_2}$, physically it means that independent measurements are performed on each qubit. Since they start each in a thermal state, by applying similar argument as above for each single qubit, one obtains, in this case, $\beta_i \langle \Delta E_i \rangle \geq 0$, that is [H]-operation.}. However, this does not necessarily mean that the post-measurement state $\rho'$, which is a mixture of them, is an entangled one. Quite remarkably, it can rather be proved on general grounds \cite{suppl} that thermodynamic efficiency is extremal at points where the post measurement state $\rho'$ is diagonal in the $\{|n\rangle\}$ basis,  that is it has no entanglement. One can check that the $\rho'$ resulting from the choice $\{|\psi^*_k\rangle \}$ above is in fact diagonal in the  $\{|n\rangle\}$ basis.

Thirdly we have found the following. Imagine to pick the measurement basis $\{|\psi_k\rangle\}$ randomly. Then, on average, the changes in the energy expectation value $\langle \Delta E_i \rangle$ is non-negative, for both $i=1,2$:
\begin{align}
\overline{\langle \Delta E_i \rangle}\geq 0\, ,
\end{align}
where the overline denotes the average over the invariant measure of $SU(4)$ (or more generally $SU(N)$ when considering a larger working substance): picking a random basis $\{|\psi_k\rangle\}$ is equivalent to picking a random unitary $U$: $|\psi_k\rangle = U |k\rangle$. That is, if choosing a random measurement basis, on average, the less useful operation, i.e. [H], is realized, independently of the choice of parameters. This means that, without any knowledge on what to do, one can only heat up everything \footnote{This result holds as well in the standard scenario where rather than interacting with a measurement apparatus, the system interacts with a classical field, and is accordingly described by a time-dependent Hamiltonian.}.
This is in fact a general result that sheds light on an interesting  facet of the second law. The general proof is presented in \cite{suppl}.

It follows that in order to realize QMC, one needs to 
know which measurement basis to use. This then opens the question of what is the probability $P_x$ that operation $x$  (with $x=[R],[E],[A],[H]$) is realized when picking a basis-change unitary $U$ randomly from the invariant $SU(4)$ measure. 
\begin{figure}[t]
\includegraphics[width=\linewidth]{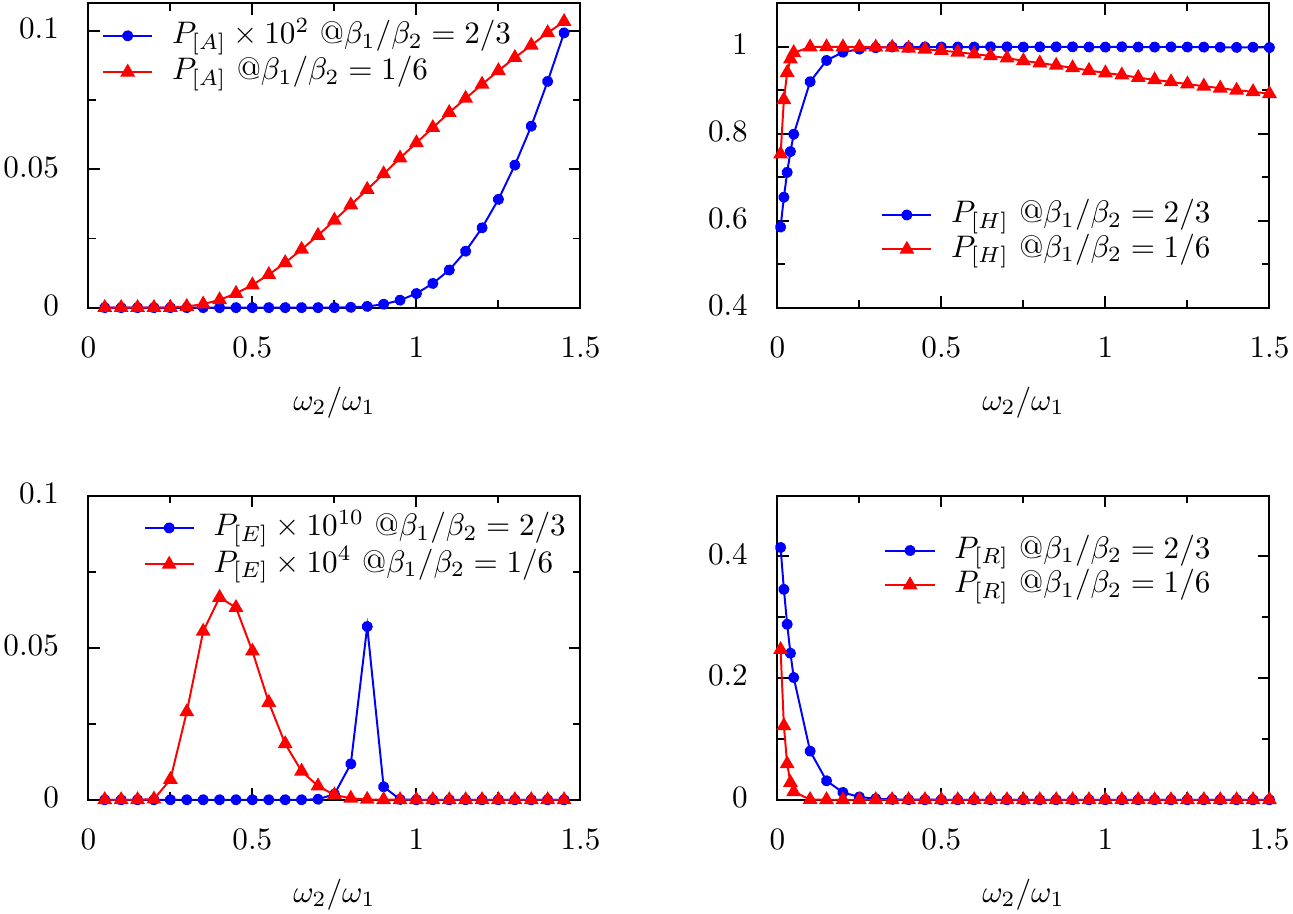}
\caption{Probability $P_x$ of the various operations $x=[A],[H], [E],[R]$ as function of level spacing ration $\omega_2/\omega_1$, at two fixed value of temperature ratios $\beta_1/\beta_2$.}
\label{fig:3}
\end{figure}
Said probability $P_x$ is given by the ratio $\mathcal{M}_x/\mathcal{M}$ of the volume $\mathcal{M}_x$ of the subset of $SU(4)$ that corresponds to [$x$]-operation rescaled by the total volume $\mathcal{M}$ of the group. Volumes are calculated with respect to the invariant (Haar) measure of the group. To quantify it we have employed the parametrization of $SU(4)$ in terms of generalised Euler angles $\boldsymbol{\alpha}=(\alpha_1, \alpha_2,\dots,  \alpha_{15})$ \cite{Tilma02JPA48}  and have performed a uniform Monte Carlo sampling of the Euler angles. We remark that such sampling is not uniform with respect to the group invariant measure $d\Omega(\boldsymbol{\alpha})=\mathcal{M}({\boldsymbol{\alpha}})d\boldsymbol{\alpha}$: To achieve uniformity over said measure each point $\boldsymbol{\alpha}$ in the sample has to be weighted with the according factor $\mathcal{M}({\boldsymbol{\alpha}})$. The results are reported in Fig. (\ref{fig:3}).

We firstly note that the Monte Carlo sampling confirms the results reported above, regarding the range of parameters associated to each operation. We also note that [H] is always the most likely operation, regardless of the parameter range.  The most surprising observation is that, while the probability $P_{[E]}$ of [E]-operation, is extremely low, the probability of [R]-operation can be very large. In fact it tends to $1/2$ from below as $\omega_2/\omega_1\rightarrow 0$. This highlights an asymmetry between the [R] and [E] operations
\footnote{Rather we observe a symmetry between the [R] and [A] operations, associated to the exchange $1\leftrightarrow2$ between hot and cold, with $P_{[R]}\rightarrow 1/2$ as  $\omega_2/\omega_1\rightarrow 0$ and $P_{[A]}\rightarrow 1/2$ as  $\omega_1/\omega_2\rightarrow 0$} having an important consequence: it shows that QMC can be made more an more robust to noise by decreasing the ratio $\omega_2/\omega_1$. This is confirmed by our numerical study showing that the region of $SU(4)$ for which QMC is realized not only grows with decreasing $\omega_2/\omega_1$ but also remains connected \cite{suppl}. Thus experimental noise on the measurement basis is not an issue with respect to implementations. In contrast, the practical feasibility of the [E]-operation is greatly hindered by the fact that $P_{[E]}$ is extremely small, hence it is extremely sensitive to experimental noise.

\emph{Considerations about the experimental realization.--} Quantum measurement cooling can be practically realized with solid-state superconducting circuitry by a suitable integration of circuit QED tools \cite{Blais04PRA69} and circuit Quantum Thermodynamics (circuit QTD) tools \cite{Pekola15NATPHYS11}. A possible design comprises two superconducting qubits coupled to an on-chip microwave line resonator  \cite{Filipp09PRL102b}. 
Using the expression $\pi_k = U\Pi_k U^\dagger$ in Eq.  (\ref{eq:rho'}) to obtain $\Phi[\rho]= \sum U\Pi_kU^\dagger \rho U\Pi_kU^\dagger$, we see that the first stroke (measurement) dynamics can be implemented by the combination of two-state manipulation and standard measurement on the  $\{|n \rangle\}$ basis, as customarily done for two-qubit tomography \cite{Filipp09PRL102b}. That is: first the gate $U^\dagger$ is applied, e.g. by driving two-photon side-band transitions  \cite{Leek09PRB79}; Then, quantum-non-demolition measurement is applied in the $\{|n\rangle\}$ basis by driving the cavity at the appropriate frequency \cite{Filipp09PRL102b}; Finally the gate $U$ is applied, e.g., by driving two-photon side-band transitions \cite{Leek09PRB79}. The qubit level spacings can be manipulated by means of local magnetic fields, and cross-resonance techniques can be used to entangle them when far detuned \cite{Paraoanu06PRB74,Rigetti10PRB81}. The output of the measurement can be inferred by reading the quadratures of the field transmitted through the resonant cavity \cite{Filipp09PRL102b}. The second stroke can be realized by inductively coupling each qubit to an on-chip resistor kept at inverse temperature $\beta_i$ \cite{Niskanen07PRB76,Karimi17QST2,Campisi17NJP19}. Heat exchanged with the resistors could be calorimetrically measured by means of fast on-chip thermometry of the resistors electron gas temperature \cite{Gasparinetti15PRAPP3,Wang18APL112}

\emph{Conclusions.--}
We have presented a genuinely quantum mechanical cooling concept, whereby the fuel is the energy exchanged with a measurement apparatus performing invasive quantum measurements. No feedback control is necessary. 
We found a number of results valid in the case of a generic two-qudit working substance: a) in order for the engine to do anything useful the measurement basis must contain entangled projectors, while, quite paradoxically, best performance is achieved when the post measurement mixture $\rho'$ is non-entangled;  b) lack of knowledge of how to operate the engine leads on average to heating up everything. Quite surprisingly in the special case of two-qubits, we have found that when choosing the measurement basis randomly, QMC can be rather likely to occur (in contrast to energy extraction), which makes its implementation robust to experimental noise. While being aware that the issue of the energetic cost of ideal projective measurement is still an actively debated fundamental problem in measurement theory \cite{Guryanova18arXiv,Busch16,Deffner16PRE94,Kammerlander16SCIREP6}, two-qubit QMC can be practically realized with superconducting circuitry by combination of circuit QED and circuit QTD (quantum thermodynamics) elements and methods.

\emph{Acknowledgements.--} The authors wish to thank Ruggero Vaia, Stefano Gherardini and Benjamin Huard for useful comments.  This research was supported in part by the National Science Foundation under Grant No. NSF PHY-1748958 and by the University of Florence in the framework of the University Strategic Project Program 2015 (project BRS00215).

\onecolumngrid

\appendix

\pagebreak

\section{Optimal measurement basis
\label{app:optimal}}

The energy change of qubit $i$ and total energy change can be written as
\begin{align}
\langle \Delta E_i \rangle = \sum_{mn} E^{(i)}_m P_{m|n}  p^0_n - \sum_n E^{(i)}_n  p^0_n \, ,\\
\langle \Delta E \rangle = \sum_{mn} E_m P_{m|n}  p^0_n - \sum_n E_n  p^0_n \, ,
\end{align}
where $ E^{(i)}_n$ and $E_n$ are the eigenvalues of $H_i$ and $H$ introduced in the main text, respectively, and $P_{m|n}$ is the probability of finding the qubits in state $m$ after the first stroke provided they first were in the state $n$. 

In the standard case, where the working substance is driven by a time dependent force generating the unitary dynamics $U$ (see main text Fig. 1 a)), the matrix $P$ is given by $P_{m|n}=|\langle m | U |n \rangle|^2$. In the case when the working substance exchanges energy with a measurement apparatus as described in the main text, the matrix $P$ has the special form $P= p^T\cdot p$, where $p$ is the bi-stochastic matrix $p_{k|n} = |\langle \psi_k|n\rangle|^2$. In both cases the matrix $P$ is bi-stochastic, that is $\sum_{m} P_{m|n} = \sum_{n} P_{m|n} =1$ \cite{Marshall11book}.

In the following we shall use the compact vector notation
\begin{align}
\langle \Delta E_i \rangle  = 
E_i^T \cdot P \cdot p^0 - E_i^T \cdot p^0 \, ,\\
\langle \Delta E \rangle = E^T \cdot P \cdot p^0 - E^T \cdot p^0 \, .
\end{align}

\subsection{[E]-range : $\beta_1/\beta_2 < \omega_2/\omega_1 <1$}
\begin{state}
\label{state:1}
In the [E]-range, among all doubly stochastic matrices $P$, the quantity 
\begin{align}
-\langle \Delta E \rangle =  E_0- E_f =  E^T \cdot p^0 - E^T \cdot P \cdot p^0
\end{align}
is maximized by the matrix $P$ that equals the matrix $\sigma_\text{S}$ representing the ``SWAP'' permutation. Among all $P$'s that realize the [E]-operation $\sigma_\text{S}$ maximizes as well the thermodynamic efficiency $\eta^{[E]}= \frac{\langle \Delta E \rangle}{\langle \Delta E_1 \rangle}$.
\end{state}
Here the explicit expression of $\sigma_\text{S}$ (all matrices below are expressed in the basis $\{|\uparrow \uparrow\rangle, |\uparrow \downarrow\rangle, |\downarrow \uparrow\rangle, |\downarrow \downarrow\rangle \}$)
\begin{align}
\sigma_\text{S}=\left(\begin{array}{cccc}1 & 0 & 0 & 0 \\0 & 0 & 1 & 0 \\0 & 1 & 0 & 0 \\0 & 0 & 0 & 1\end{array}\right)\, .
\end{align}

The permutation $\sigma_\text{S}$ can be realized by means of a unitary gate, in fact the SWAP gate  $U_\text{S}$, $[\sigma_\text{S}]_{mn}=|\langle m| U_\text{S} |n \rangle|^2$. That is the statement says that driving the two-qubit compound by means of a driving force realizing a a SWAP gate maximizes energy extraction among all possible unitaries. This is a result that has been previously observed numerically \cite{Campisi15NJP17}. However it cannot be realized by a projective measurement channel, for which the transition matrix takes on the special form $P = p^T \cdot p$, with $p$ itself a bi-stochastic matrix. That is there exist no doubly stochastic matrix $p$ such that $\sigma_\text{S} = p^T \cdot p$. 

\begin{state}
In the [E]-range, among all doubly stochastic matrices $P$, the quantity 
\begin{align}
-\langle \Delta E \rangle =  E_0- E_f = E_0 - E^T \cdot P^T \cdot P \cdot p^0
\end{align}
is maximized by the matrix  $\sigma_{SI}=(\sigma_I+\sigma_\text{S})/2$ with $\sigma_I$ the $4 \times 4$ identity matrix.
Among all $P$'s that realize the [E]-operation $\sigma_{SI}$ maximizes as well the thermodynamic efficiency $\eta^{[E]}= \frac{\langle \Delta E \rangle}{\langle \Delta E_1 \rangle}$.
\end{state}
Here the explicit expression of $\sigma_{SI}$
\begin{align}
\sigma_{SI}=\left(\begin{array}{cccc}1 & 0 & 0 & 0 \\0 & \nicefrac{1}{2} & \nicefrac{1}{2} & 0 \\0 & \nicefrac{1}{2} & \nicefrac{1}{2} & 0 \\0 & 0 & 0 & 1\end{array}\right)\, .
\end{align}
It can be realized (modulo unimportant relative and global phases) with the basis change unitary
\begin{align}
U_{*}=\left(\begin{array}{cccc}1 & 0 & 0 & 0 \\0 & \nicefrac{1}{\sqrt{2}} &  \nicefrac{1}{\sqrt{2}} & 0 \\0 &  \nicefrac{1}{\sqrt{2}} & -\nicefrac{1}{\sqrt{2}} & 0 \\0 & 0 & 0 & 1\end{array}\right)\, ,
\label{eq:sqrt-SWAP}
\end{align}
giving the optimal basis presented in the main text via the relation $|\psi_k^*\rangle = U_*|k\rangle$.

\subsubsection{Proof of Statement 1}

By Birkoff/Von Neumann theorem \cite{Marshall11book} a doubly stochastic matrix $P$ can be written as convex combination of permutation matrices $\sigma_\alpha$, $\alpha=1,...4!$
\begin{align}
P =  \sum_{i=1}^{n!} \mu_i \sigma_i,  \quad 
\sum_i \mu_i = 1,  \qquad 0\leq \mu_i \leq 1\, .
\end{align} 
We need not list here all the permutation matrices, it suffices, for now, to say that we label them in such a way that 
$\sigma_1=\sigma_I=\mathbb{1}$, $\sigma_2=\sigma_S$. The symbol $\sigma_i$ appearing here should not be confused with the Pauli matrices appearing in the main text.

In vector notation we have
\begin{align}
E_f=\sum_{i} \mu_i E^T \cdot \sigma_{i} \cdot p^0= \sum_{i} \mu_i u_i\, ,
\end{align}  
where
$
u_i= E^T \cdot \sigma_{i} \cdot p^0
$
is a real number. We want to minimize $E_f$. By the method of Lagrange multipliers we impose $0=\partial (E_f-\gamma \sum \mu_i)/\partial \mu_j= u_j-\gamma$ which, generally, has no solution, because the $u_j$ are generally not all equal. This implies that the minimum of $E_f$ is on the boundary of the polytope defined by the equation $\sum_i \mu_i=1$. The minimum occurs trivially when the smallest among all $u_i$ has largest weight. By direct inspection we find that in the [E]-regime $u_2$ is the smallest among all $u_j$'s
\begin{align}
u_2 \leq u_i  \qquad \forall i\, .
\end{align}
That is the minimum of $\langle \Delta E\rangle $ occurs for $\mu_i=0, \forall i\neq 2; \mu_2 =1$. The corresponding permutation matrix is
$\sigma_2=\sigma_\text{S}$. So the minimum of $E_f$ is reached for the bi-stocastic matrix $P=\sigma_\text{S}$. 

In the following we shall show that $\sigma_S$ maximizes efficiency as well.
Introducing the notation $u^{(1)}_j= E^{(1)T}\cdot \sigma_j \cdot p^0$, $E_0^{(1)}= E^{(1)T}\cdot p^0=u^1_1$ we have
\begin{align}
\eta^{[E]}&=\frac{\langle\Delta E\rangle}{\langle\Delta E_1\rangle}\\
&=\frac{\sum_i \mu_i(u_i-E_0)}{\sum_i \mu_i(u^{(1)}_i-E_0^{(1)}) } \\
&= \frac{ \mu_2(u_2-E_0) + \sum_{i\neq 2} \mu_i(u_i-E_0)} {\mu_2(u_2^{(1)}-E_0^{(1)})+\sum_{i\neq 2}\mu_i(u^{(1)}_i-E_0^{(1)})} \\
&=\frac{ \mu_2(u_2-E_0)+A}{\mu_2(u_2^{(1)}-E_0^{(1)})+B}\, ,
\end{align}
where
\begin{align}
A&=\sum_{i\neq 2} \mu_i(u_i-E_0)= \sum_{i\neq 1,2} \mu_i(u_i-E_0)\, ,\\
B&=\sum_{i\neq 2}\mu_i(u^{(1)}_i-E_0^{(1)})= \sum_{i\neq 1,2}\mu_i(u^{(1)}_i-E_0^{(1)})\, .
\end{align}
The last equalities follow from $u_1=E_0$ and $u_1^1=E_0^{(1)}$. We remark, that we are now restricting the minimisation to the subset of $P$'s that realized the [E]-operation, that is for which $\langle\Delta E_1\rangle=\mu_2(u_2^{(1)}-E_0^{(1)})+B\leq0$.
Then
\begin{align}
A-B&=\sum_{i\neq 1,2} \mu_i(u_i-E_0-u^{(1)}_i+E_0^{(1)})=\sum_{i\neq 1,2} \mu_i(u_i^{(2)}-E_0^{(2)})\, ,
\end{align} 
where $u_i^{(2)}= E^{(2)T}\cdot \sigma_j \cdot p^0 = u-u^{(1)}_i$, $E_0^{(2)}= E^{(2)T}\cdot p^0=E_0-E_0^{(1)}$.
By direct inspection we see that in the [E]-region it is $u_i^{(2)}-E_0^{(2)}\geq 0$ for all $i\neq 1,2$, hence $A\geq B$.  Furthermore,
in the [E]-region it is:
\begin{align} 
\mu_2(u_2-E_0)\leq 0 \qquad \mu_2(u_2^{(1)}-E_0^{(1)})\leq 0\, .
\end{align}
We want to prove that 
\begin{align}
\eta^{[E]}=\frac{ \mu_2(u_2-E_0)+A}{\mu_2(u_2^{(1)}-E_0^{(1)})+B}\leq\frac{u_2-E_0}{u_2^{(1)}-E_0^{(1)}}\, ,
\label{eq:Eefficiency}
\end{align}
or, equivalently,
\begin{align}
(\mu_2(u_2-E_0)+A)(u_2^{(1)}-E_0^{(1)})\leq (\mu_2(u_2^{(1)}-E_0^{(1)})+B)(u_2-E_0)\, ,
\end{align}
where we used the condition of [E]-operation,  $\langle \Delta E_1 \rangle = \mu_2(u_2^{(1)}-E_0^{(1)})+B \leq 0$.
By simple algebraic manipulation it follows 
\begin{align}
A(u_2^{(1)}-E_0^{(1)}) \leq B (u_2-E_0)\, ,
\end{align}
that is,
\begin{align}
\frac{A}{B}\geq \frac{u_2-E_0}{u_2^{(1)}-E_0^{(1)}}\, .
\label{eq:A/B}
\end{align}
We note that the ratio $(u_2-E_0)/(u_2^{(1)}-E_0^{(1)})$ is the thermodynamic efficiency relative to $P=\sigma_S$, which is equal to $1-\omega_2/\omega_1$ \cite{Campisi15NJP17}. Since $A/B\geq 1$ then $A/B\geq 1-\omega_2/\omega_1$, that is Eq. (\ref{eq:A/B}) is obeyed. Then Eq. (\ref{eq:Eefficiency}) follows, which concludes the proof.

\subsubsection{Proof of Statement 2}

We have, in vector notation
\begin{align}
E_f = \sum_{i j } \mu_i \mu_j  E^T \cdot \sigma_i^T \cdot \sigma_j \cdot p^0=  \sum_{i j } \mu_i \mu_j u_{i,j}\, ,
\end{align}
where
\begin{align}
u_{i,j}=  E^T \cdot \sigma_i^T \sigma_j \cdot p^0
\end{align}
is a real number. 
Note that only the symmetric part $u^S=(u+u^T)/2$ of the $n! \times n!$ matrix $u$ plays a role in the sum.
Note also that $u_{i,i}^S=u_{i,i}=E_0$ where $E_0=E^T \cdot  p^0$ is the initial energy expectation, because for a permutation matrix it is $\sigma_i \sigma_i^T=\mathbb{1}$.

Let $u_{0}$ be the smallest among the $u_{i,j}^S$'s
\begin{align}
u_{0}\leq u_{i,j}^S \qquad \text{for all}\qquad i,j\, .
\label{eq:barU}
\end{align}
By direct inspection we find that this minimum is obtained for  $\sigma_i^T\sigma_j= \sigma_j^T\sigma_i=\sigma_S=\sigma_2$ that is it corresponds to the labels $(i,j)=(1,2)$ or $(i,j)=(2,1)$.
By direct inspection we also see that  in the [E]-range  ($\beta_1/\beta_2 \leq \omega_2/\omega_1\leq 1$), it is 
\begin{align}
E_0 \leq  u_{i,j}^S \qquad \text{for} \qquad (i,j) \neq (1,2); (i,j) \neq (2,1)\, ,
\label{eq:E0}
\end{align}
where $E_0$ is the initial energy. We have
\begin{align}
E_f &=  \sum_{i\neq j}   \mu_i \mu_j u_{i,j}^S + \sum_i \mu_i^{(2)} E_0 \\
&= \sum_{i\neq j}  u_{i,j}^S \mu_i \mu_j + (1-\sum_{i\neq j}  \mu_i \mu_j) E_0 \\
&= 2 \mu_2 \mu_1 u_{0}  + \sum'_{i\neq j}   \mu_i \mu_j u_{i,j}^S + (1-2\mu_2 \mu_1 -\sum'_{i\neq j}  \mu_i \mu_j) E_0 \\
&= 2 \mu_2 \mu_1 (u_{0}-E_0)  + E_0 + \sum'_{i\neq j}   \mu_i \mu_j (u_{i,j}^S-E_0) \, ,
\end{align}
where $\sum'$ is restricted to $(i,j) \neq (1,2),(2,1)$. Now it is 
\begin{align}
\mu_2 \mu_1= \mu_2(1-\mu_2-\sum_{i\neq 1,2}\mu_i) \leq \mu_2(1-\mu_2) \leq 1/4\, ,
\end{align}
and $u_{0}-E_0\leq 0$, hence
\begin{align}
E_f & \geq \frac{1}{2} (u_{0}-E_0)  + E_0 + \sum'_{i\neq j}   \mu_i \mu_j (u_{i,j}^S-E_0) \geq \frac{1}{2} (u_{0}+ E_0)\, .
\end{align}
This bound can be saturated with the choice $\mu_1=\mu_2=1/2$ and $\mu_i=0$ for $i\neq 1,2$. Therefore $(u_{0}+ E_0)/2$ is the minimum of $E_f$.

The choice $\mu_1=\mu_2=1/2$ corresponds to the choice $P=(\sigma_1+\sigma_2)/2=(\sigma_I+\sigma_\text{S})/2$, hence $q=P^T\cdot P = (\sigma_I+\sigma_\text{S})/2$, which proves the first part of the statement. By explicit calculation we find that the corresponding energy extraction efficiency is $\eta= 1-\omega_2/\omega_1$, which is maximal among all doubly stochastic matrices (see Statement 1), hence as well among all matrices of the form $P^T \cdot P$, (with $P$ itself doubly stochastic), which are themselves doubly stochastic, which proves the second part of the statement.

\subsection{[R]-range}

\begin{state}\label{state:3}
In the [R]-range, among all doubly stochastic matrices $P$, the quantity 
\begin{align}
- \langle \Delta E_2 \rangle = -E^{(2)}_f +  E^{(2)}_0= - E^{(2) T} \cdot P \cdot p^0 +  E^{(2) T} \cdot p^0
\end{align}
is maximized by the matrix $\sigma_\text{S}$ representing the ``SWAP'' permutation. Among all $P$'s that realize the [R]-operation $\sigma_\text{S}$ maximizes as well the refrigeration efficiency $\eta^{[R]}= -\frac{\langle \Delta E_2 \rangle}{\langle \Delta E \rangle}$.
\end{state}

\begin{state}
In the [R]-range, among all doubly stochastic matrices $P$, the quantity 
\begin{align}
- \langle \Delta E_2 \rangle = -E^{(2)}_f +  E^{(2)}_0= - E^{(2) T} \cdot P^T \cdot P \cdot p^0 +  E^{(2) T} \cdot p^0
\end{align}
is maximized by the matrix  $\sigma_{SI}=(\sigma_I+\sigma_\text{S})/2$ with $\sigma_I$ the $4 \times 4$ identity matrix.
Among all $P$'s that realize the [E]-operation  $\sigma_{SI}$ maximizes as well the thermodynamic efficiency $\eta^{[R]}= -\frac{\langle \Delta E_2 \rangle}{\langle \Delta E \rangle}$.
\end{state}

\subsubsection{Proof of Statement 3}
The proof follows the same idea of that of Statement \ref{state:1}, with $E_f$ being replaced by
\begin{align}
E_f^{(2)}=\sum_{i} \mu_i E^{(2)T} \cdot \sigma^{i} \cdot p^0= \sum_{i} \mu_i u_i^{(2)} \, .
\end{align}
The main difference is now that there are many $k$'s for which  $u^{(2)}_k$ gets the smallest value. With our labelling they correspond to $k=2,5,9$,  where the corresponding permutation matrices are 
\begin{align}
\sigma_5=\begin{pmatrix}
1 &0 &0 &0\\
0 &0 &0 &1\\
0 &1 &0 &0\\
0 &0 &1 &0\\
\end{pmatrix}
,\qquad
\sigma_9=\begin{pmatrix}
0 &1 &0 &0\\
0 &0 &1 &0\\
1 &0 &0 &0\\
0 &0 &0 &1\\
\end{pmatrix}
.
\end{align}
So every combination of the type
\begin{align}
P=\mu_1\sigma_2+\mu_5\sigma_5+\mu_9\sigma_9
\end{align}
maximizes the amount of energy extracted from the cold bath $-\langle\Delta E_2\rangle $. In particular the choice $P= \sigma_2=\sigma_S$ maximizes the energy extraction from the cold bath, thus proving the first part of the statement.

Refrigeration efficiency is given by
\begin{align}
\eta^{[R]}=-\frac{\langle\Delta E_2\rangle}{\langle\Delta E\rangle}\, .
\end{align}
Now in the [R] region we have
\begin{align}
-\langle\Delta E_2 \rangle &\leq -\epsilon(u_2^{(2)}-E_0^{(2)})\, ,\\
\langle\Delta E\rangle &\geq \epsilon(u_2-E_0)\geq 0\, .
\end{align}
So we have an upper bound for the efficiency
\begin{align}
\eta^{[R]}\leq -\frac{\epsilon(u_2^{(2)}-E_0^{(2)})}{\epsilon(u_2-E_0)}= -\frac{(u_2^{(2)}-E_0^{(2)})}{(u_2-E_0)}\, .
\end{align}
This bound can be saturated with the choice $\mu_2=1$ and $\mu_i=0$ for all $i\neq 2$.
Summing up we have shown that the SWAP bistocastic $P=\sigma_s$ simultaneously maximizes $-\langle\Delta E_2\rangle$ and the refrigeration efficiency $\eta^{[R]}$. The corresponding maximal value of efficiency reads $\eta^{[R]}= 1/(\omega_1/\omega_2-1)$.

\subsubsection{Proof of Statement 4}

By virtue of Birkhoff/von Neumann theorem, we have
\begin{align}
E^{(2)}_f = \sum_{i j } \mu_i \mu_j  E^{(2)T}\sigma_i^T \sigma_j p^0 = \sum_{i j } \mu_i \mu_j A_{i,j}  = \sum_{i j } \mu_i  \mu_jA_{i,j}^S\, , 
\end{align}
where $
E^{(2)T}\sigma_i^T \sigma_j p^0 = A_{i,j}
$ is a real number.
Only the symmetric part $A^S=(A+A^T)/2$ of the $n! \times n!$ matrix $A$ plays a role in the sum.
Note that $A_{i,i}^S=A_{i,i}=E_2^0$ where $E_2^0=E^{(2)T} p^0$ is the initial energy expectation of qubit 2.
Let $A^*$ be the smallest among the $A_{i,j}^S$'s:
\begin{align}
A^*\leq A_{i,j}^S \qquad \text{for all}\qquad i,j\, .
\label{eq:barA}
\end{align}
By direct inspection we find
\begin{align}
A^*=A_{1,2}=A_{1,5}=A_{1,9} \, ;
\end{align}
we also notice that in the region  $\omega_2/\omega_1\leq\beta_1/\beta_2\leq 1$, it is
\begin{align}
A^*\leq\ A_0\leq A_{i,j}^S\qquad \text{for all}\qquad (i,j) \neq (1,2);(1,5);(1,9);(2,1);(5,1);(9,1)\, .
\end{align}
We have
\begin{align}
E_{f}^2&= \sum_{i j } \mu_i\mu_jA_{i,j}^S\\
&=\sum_{i\neq j}\mu_i\mu_jA_{i,j}^S + \sum_{i}\mu_i^{(2)}A_0\\
&=\sum_{i\neq j}\mu_i\mu_jA_{i,j}^S + (1-\sum_{i\neq j}\mu_i\mu_j)A_0\\
&=2(\mu_1\mu_2+\mu_1\mu_5+\mu_1\mu_9)A^* + \sum'_{i\neq j}\mu_i\mu_jA_{i,j}^S + (1-2(\mu_1\mu_2+\mu_1\mu_5+\mu_1\mu_9)-\sum'_{i\neq j}\mu_i\mu_j)A_0\\
&=2\mu_1(\mu_2+\mu_5+\mu_9)(A^*-A_0) + A_0 + \sum'_{i\neq j}\mu_i\mu_j(A_{i,j}^S-A_0)\, ,
\end{align} 
where $\sum'$ is restricted to the couple of $i,j$ such that $(i,j)\neq (1,2);(1,5);(1,9),(2,1);(5,1);(9,1)$. Now it is
\begin{align}
\mu_1(\mu_2+\mu_5+\mu_9)=\mu_1(1-\mu_1-\sum_{i\neq 1,2,5,9}\mu_i) \leq \mu_1(1-\mu_1) \leq 1/4\, ,
\end{align} 
and $A^*\leq A_0$, hence
\begin{align}
E_{f}^2 \geq \frac{1}{2}(A^*-A_0)+A_0+\sum'_{i\neq j}\mu_i\mu_j(A_{i,j}^S-A_0) \geq \frac{1}{2}(A^*+A_0)\, .
\end{align}
This bound can be saturated in three different ways:
\begin{itemize}
\item $\mu_1=\mu_2=\frac{1}{2}$ and $\mu_i=0$ $\forall i\neq 1,2$ 
\item $\mu_1=\mu_5=\frac{1}{2}$ and $\mu_i=0$ $\forall i\neq 1,5$ 
\item $\mu_1=\mu_9=\frac{1}{2}$ and $\mu_i=0$ $\forall i\neq 1,9$ 
\end{itemize} 

The choice $\mu_1=\mu_2=1/2$ corresponds to the choice $P=(\sigma_I+\sigma_\text{S})/2$, hence $P^T\cdot P = (\sigma_I+\sigma_\text{S})/2$ which proves the first part of the statement. By explicit calculation we find that the corresponding energy extraction efficiency is $\eta^{[R]}= 1/(\omega_1/\omega_2-1)$, which is maximal among all doubly stochastic matrices (see Statement \ref{state:3}), hence as well among all matrices of the form $P^T \cdot P$, (with $P$ itself doubly stochastic), which are themselves doubly stochastic.

Summing up $(\sigma_I+\sigma_\text{S})/2$ simultaneously maximizes $- \langle \Delta E_2\rangle$ and the efficiency $\eta^{[R]}= -\langle \Delta E_2 \rangle/\langle \Delta E \rangle$.

\section{The two-qudit case}
In this case the analysis gets considerably more involved. Let us consider the problem of minimising
\begin{align}
E_f = \sum_{i j } \mu_i \mu_j  E^T \cdot \sigma_i^T \cdot \sigma_j \cdot p^0 =  \sum_{i j } \mu_i \mu_j u_{i,j}\, ,
\end{align}
where, as noted before 
\begin{align}
u_{i,j}=  E^T \cdot \sigma_i^T \sigma_j \cdot p^0
\end{align}
is a real number, and only the symmetric part $u^S=(u+u^T)/2$ of the $n! \times n!$ matrix $u$ plays a role in the sum:
\begin{align}
u_{i,j}^S=  E^T \cdot \frac{(\sigma_i^T \sigma_j + \sigma_j^T \sigma_i)}{2}\cdot p^0
\end{align}
Let $u_{0}^S$ be the smallest among the $u_{i,j}^S$'s
\begin{align}
u_{0}^S\leq u_{i,j}^S \qquad \text{for all}\qquad i,j\, .
\label{eq:barU}
\end{align}
Let $u_{1}^S$ be the second smallest among the $u_{i,j}^S$'s.
In general $u_1^S$ can be smaller than $E_0$ and there might be more $u$'s between $u_1^S$ and $E_0$
\begin{align}
u_{0}^S\leq u_1^S \leq \dots \leq E_0 \leq \dots
\label{eq:barU}
\end{align}
We assume there is only $u_1^S$ between $u_0^S$  and $E_0$ (the treatment of the more general case proceeds straightforwardly), that is we assume the ordering
\begin{align}
u_{0}^S\leq u_1^S \leq E_0 \leq \dots
\label{eq:barU}
\end{align}
For sake of simplicity we assign the labels $1,2$ to the couple of permutations associated to $u_0^S$: $u_0^S=u_{1,2}^S$.
This can always be done because if $i^*,j^*$ are  such that $u_{i^*,j^*}^S=u_0^S$, since $\sigma_{i^*}^T \sigma_{j^*}$ is a permutation, there is a $k^*$ such that $\sigma_{i^*}^T \sigma_{j^*}=\sigma_{k^*}=\sigma_1^T \sigma_{k^*}$, where $\sigma_1$ is the identity. Hence $\sigma_{i^*}^T \sigma_{j^*} + \sigma_{j^*}^T \sigma_{i^*} =\sigma_1^T \sigma_{k^*}+  \sigma_{k^*}^T \sigma_1$, and $u_{i^*,j^*}^S=u_0= u_{k^*,1}^S$, that is we can replace $i^*,j^*$ with $k^*,1$. Similarly we assign the labels $1,3$ to the couple of permutations associated to $u_1^S$: $u_1^S=u_{1,3}^S$. 

We have
\begin{align}
E_f &=  \sum_{i\neq j}   \mu_i \mu_j u_{i,j}^S + \sum_i \mu_i^2 E_0 \\
&= \sum_{i\neq j}  u_{i,j}^S \mu_i \mu_j + (1-\sum_{i\neq j}  \mu_i \mu_j) E_0 \\
&= 2 \mu_2 \mu_1 u_{0}^S  + 2 \mu_1 \mu_3 u_1^S + \sum'_{i\neq j}   \mu_i \mu_j u_{i,j}^S + E_0 -(2\mu_2 \mu_1 +2\mu_1\mu_3+ \sum'_{i\neq j}  \mu_i \mu_j) E_0 \\
&= 2 \mu_2 \mu_1 (u_{0}^S-E_0) + 2 \mu_1 \mu_3 (u_{1}^S-E_0)  + E_0 + \sum'_{i\neq j}   \mu_i \mu_j (u_{i,j}^S-E_0) \, \\
& \geq 2 \mu_2 \mu_1 (u_{0}^S-E_0) + 2 \mu_1 \mu_3 (u_{1}^S-E_0)  + E_0 \\
& \geq 2 \mu_1(\mu_2+\mu_3) (u_{0}^S-E_0)  + E_0 \label{eq:Efbound-general} \\
& \geq  (u_{0}^S+E_0)/2
\end{align}
where $\sum'$ is restricted to $(i,j) \neq (1,2),(2,1),(1,3),(3,1)$. The last step follows from the inequality
$\mu_1(\mu_2+\mu_3) \leq 1/4\, $. This bound can be saturated with the choice $\mu_1=\mu_2=1/2$ and $\mu_i=0$ for $i\neq 1,2$. Therefore $(u_{0}^S+ E_0)/2$ is the minimum of $E_f$. The choice $\mu_1=\mu_2=1/2$ corresponds to the choice $P=(\sigma_1+\sigma_2)/2$.  The following two cases may occur, but, as we shall see below, only the first one is physically relevant.

\subsubsection{Case 1: $\sigma_2$ is symmetric}
In this case $\sigma_2$ represents a permutation composed of disjoint cycles of length 2. Here is a representative example of such permutation in the two-qutrit case where the 4th state takes the place of the 6th and viceversa while the 3rd  takes the place of the 8th and viceversa:
\begin{align}
\left(\begin{array}{ccccccccc}
1 & 0 & 0 & 0 & 0 & 0 & 0 & 0 & 0 \\
0 & 1 & 0 & 0 & 0 & 0 & 0 & 0 & 0 \\
0 & 0 & 0 & 0 & 0 & 0 & 0 & 1 & 0 \\
0 & 0 & 0 & 0 & 0 & 1 & 0 & 0 & 0 \\
0 & 0 & 0 & 0 & 1 & 0 & 0 & 0 & 0 \\
0 & 0 & 0 & 1 & 0 & 0 & 0 & 0 & 0 \\
0 & 0 & 0 & 0 & 0 & 0 & 1 & 0 & 0 \\
0 & 0 & 1 & 0 & 0 & 0 & 0 & 0 & 0 \\
0 & 0 & 0 & 0 & 0 & 0 & 0 & 0 & 1
\end{array}\right)
\end{align}
The according $P=(\sigma_1+\sigma_2)/2$ is
\begin{align}
\left(\begin{array}{ccccccccc}
1 & 0 & 0 & 0 & 0 & 0 & 0 & 0 & 0 \\
0 & 1 & 0 & 0 & 0 & 0 & 0 & 0 & 0 \\
0 & 0 & 1/2 & 0 & 0 & 0 & 0 & 1/2 & 0 \\
0 & 0 & 0 & 1/2 & 0 & 1/2 & 0 & 0 & 0 \\
0 & 0 & 0 & 0 & 1 & 0 & 0 & 0 & 0 \\
0 & 0 & 0 & 1/2 & 0 & 1/2 & 0 & 0 & 0 \\
0 & 0 & 0 & 0 & 0 & 0 & 1 & 0 & 0 \\
0 & 0 & 1/2 & 0 & 0 & 0 & 0 & 1/2 & 0 \\
0 & 0 & 0 & 0 & 0 & 0 & 0 & 0 & 1
\end{array}\right)
\end{align}
Such matrices involving only $2\times2$ blocks with all entries being $1/2$ can be generated with the corresponding basis-change  that includes only factorized states $|a,a\rangle$ and states of the type $(|a,b\rangle \pm |c,d\rangle)/\sqrt{2}$ (where not necessarily $a,b,c,d$ are all different from each other) as stated in the main text. 

\subsubsection{Case 2: $\sigma_2$ is not symmetric}

In this case the permutation represented by $\sigma_2$ is composed of cycles of length larger than 2. Here is a representative of a not-symmetric permutation in the two-qutrit case, where the 2nd state takes the place of the 3th, the 3rd state takes the place of the 4th, and the 4th takes the place of the 2nd:
\begin{align}
\left(\begin{array}{ccccccccc}1 & 0 & 0 & 0 & 0 & 0 & 0 & 0 & 0 \\0 & 0 & 0 & 1 & 0 & 0 & 0 & 0 & 0 \\0 & 1 & 0 & 0 & 0 & 0 & 0 & 0 & 0 \\0 & 0 & 1 & 0 & 0 & 0 & 0 & 0 & 0 \\0 & 0 & 0 & 0 & 1 & 0 & 0 & 0 & 0 \\0 & 0 & 0 & 0 & 0 & 1 & 0 & 0 & 0 \\0 & 0 & 0 & 0 & 0 & 0 & 1 & 0 & 0 \\0 & 0 & 0 & 0 & 0 & 0 & 0 & 1 & 0 \\0 & 0 & 0 & 0 & 0 & 0 & 0 & 0 & 1\end{array}\right)
\end{align}
The according $P=(\sigma_1+\sigma_2)/2$ is
\begin{align}
\left(
\begin{array}{ccccccccc}
 1 & 0 & 0 & 0 & 0 & 0 & 0 & 0 & 0
   \\
 0 & \nicefrac{1}{2} & 0 & \nicefrac{1}{2} &
   0 & 0 & 0 & 0 & 0 \\
 0 & \nicefrac{1}{2} & \nicefrac{1}{2} & 0 &
   0 & 0 & 0 & 0 & 0 \\
 0 & 0 & \nicefrac{1}{2} & \nicefrac{1}{2} &
   0 & 0 & 0 & 0 & 0 \\
 0 & 0 & 0 & 0 & 1 & 0 & 0 & 0 & 0
   \\
 0 & 0 & 0 & 0 & 0 & 1 & 0 & 0 & 0
   \\
 0 & 0 & 0 & 0 & 0 & 0 & 1 & 0 & 0
   \\
 0 & 0 & 0 & 0 & 0 & 0 & 0 & 1 & 0
   \\
 0 & 0 & 0 & 0 & 0 & 0 & 0 & 0 & 1
   \\
\end{array}
\right)
\label{eq:p}
\end{align}
In this case, it is not guaranteed that a basis change unitary $U$ exists such that $P_{i,j}=|\langle i|U|j\rangle |^2$, that is it is not guaranteed that $P$ is a so-called unitary stochastic matrix, in short unistochastic \cite{Marshall11book}. In fact,  it cannot be unistochastic. As shown in Ref. \cite{Yik-Hoi91LINALGAPP150} the following holds:
\begin{theo}
Any convex combination of permutation matrices which is unistochastic, is such that it is composed of permutations that are pairwise complementary
\end{theo}
where two $n \times n$ matrices $A$ and $B$ are said to be complementary if, for any $1\leq i,j,h,k\leq n$, $A_{ij}=A_{hk}=B_{ik}=1$ implies $B_{hj}=1$ \footnote{We warn the reader that  Ref. \cite{Yik-Hoi91LINALGAPP150} uses the therm ``ortostochastic'' to designate what here we refer to as ``unistochastic''. See also \cite{Chterental08LAA428} for the same warning.}. Since any permutation that is complementary to the identity is symmetric \cite{Yik-Hoi91LINALGAPP150}, this implies that $P=(\sigma_1+\sigma_2)/2$ with a non-symmetric $\sigma_2$ cannot, in fact, be unistochastic.

In the light of the above theorem we see that, when searching for the minimum of $E_f$ on the set of physically relevant unistochastic matrices, one should impose the constraint that the Birkhoff/Von Neumann expansion $P=\sum \mu_i \sigma_i$, contains only pairwise complementary permutations. The latter either includes the identity, in which case only symmetric matrices are allowed and only Case 1 may occur, or it does not include it. By looking at Eq. (\ref{eq:Efbound-general}) we see that not including the identity (i.e. imposing $\mu_1=0$) immediately lifts up the minimal value that the expansion may reach, which now becomes $E_0$, thus preventing not only the possibility of reaching the true minimum that may well be  below $E_0$, but also that the regime where the device actually works as a heat engine may be reached. It follows that this Case 2 is excluded on 
physical grounds.

\subsection{Numerical analysis}

The above theory shows that the minimum of $E_f$ can always be reached by a basis-change  that includes only factorized states $|a,a\rangle$ and Bell-type states $(|a,b\rangle \pm |c,d\rangle)/\sqrt{2}$, as discussed in the main text. This was further corroborated by a numerical analysis.

We have numerically inspected whether $\sigma_2$ is symmetric for the case of two three-level systems with equally spaced levels, with resonance frequencies $\omega_1$ and $\omega_2$. This amounted to finding which of the $9!=362880$ permutations minimizes the quantity $E_f=E^T \cdot {(\sigma_k^T + \sigma_k)}/{2}\cdot p^0$. This was done for $\beta_1=1$  
$\beta_2=\lbrace 1,2,3,4\rbrace$, and $(\omega_1,\omega_2)$ ranging in the region $(0,4]$ with a resolution of $0.25$, for a total of about $1000$ trials. In many cases the minimum was achieved by several permutations, but always there was at least one symmetric permutation among them. So Case 1 always occurred, while Case 2 often occurred, but always in concomitance with Case 1, thus corroborating that the minimum can always be achieved by a basis change that includes only factorized states $|a,a\rangle$ and states of the form $(|a,b\rangle \pm |c,d\rangle)/\sqrt{2}$.

Few more trials were performed in the 4-level case. In this case we looked for the minimum of $E_f = \Tr \sum H U\Pi_kU^\dagger \rho U\Pi_kU^\dagger$ over all possible $U's$ by using MATLAB's ``fmincon'' function, as spanning the whole Birkhoff polytope generated by the $16! \simeq 10^{13}$ permutations is numerically prohibitive. In this case, in accordance with the theory above, we always found the minimum being reached by a basis that includes only factorized states $|a,a\rangle$ and states of the form $(|a,b\rangle \pm |c,d\rangle)/\sqrt{2}$.

\section{Condition for optimal energy extraction by measurement\label{app:condition}}
n the following we shall prove that if the post-measurement state $\rho'$ commutes with the Hamiltonian $H$, then the total energy injection $\langle \Delta E \rangle$ is at an extremum, namely we shall prove  $[\rho',H]=0 \implies \delta \langle \Delta E \rangle=0$, where the variation is taken with respect to a variation of the measurement basis. We have
\begin{align}
\rho'= \Phi[\rho] = \sum_{k}   \pi_k \rho \pi_k  = \sum_k U \Pi_k U^\dagger \rho U \Pi_k U^\dagger\, ,
\end{align} 
where $\Pi_k$ are the $H$ eigen-projectors and $U$ is the basis-change unitary. The variation $\langle \Delta E \rangle$ is null when the variation of the final energy 
\begin{align}
E_f=\mbox{Tr}\left[\Phi[\rho]H\right]=\mbox{Tr}\left[ \sum_k U \Pi_k U^\dagger \rho U \Pi_kU^\dagger H\right]
\end{align}
is null. Following \cite{Allahverdyan04EPL67} we parametrise the variation $\delta U$ of $U$ as $\delta U=XU$ where $X$ is some anti-Hermitian operator. Hence we find
\begin{align}
\delta E_f&=\mbox{Tr}\left[\sum_k \left(\delta U\Pi_kU^\dagger \rho U\Pi_k U^\dagger+ U\Pi_k\delta U^\dagger \rho U\Pi_kU^\dagger + U\Pi_kU^\dagger \rho \delta U\Pi_kU^\dagger+ U\Pi_kU^\dagger \rho U\Pi_k\delta U^\dagger\right) H\right]\\
&=\mbox{Tr}\left[\sum_k \left(X U\Pi_kU^\dagger \rho U\Pi_k U^\dagger- U\Pi_k U^\dagger X\rho U\Pi_kU^\dagger + U\Pi_kU^\dagger \rho X U\Pi_kU^\dagger- U\Pi_kU^\dagger \rho U\Pi_kU^\dagger X\right) H\right]\\
&=\mbox{Tr}\left[\left(X\Phi[\rho]-\Phi[\rho]X\right)\right]+\mbox{Tr}\left[\left(\rho X-X\rho\right)H\right]\, ,
\end{align}
where in the last line we have used the fact that $\sum_kU\Pi_kU^\dagger=\mathbb{1}$. Now using the ciclicity of the trace we obtain
\begin{align}
\delta E_f&=\mbox{Tr}\left[X \left(\Phi[\rho]H-H\Phi[\rho]\right)\right]+\mbox{Tr}\left[X\left(H\rho-\rho H\right)\right]=\mbox{Tr}\left(X\left[\Phi[\rho],H\right]\right)+\mbox{Tr}\left(X\left[H,\rho\right]\right)\, .
\end{align}
Since the initial state $\rho$ is thermal we have that $\left[H,\rho\right]=0$, therefore
\begin{align}
\delta E_f=\mbox{Tr}\left(X\left[\Phi[\rho],H\right]\right)\, ,
\end{align}
hence $[\rho',H]=0 \implies \delta \langle \Delta E \rangle=0$.

Similarly, $[\rho',H_i]=0 \implies \delta \langle \Delta E_i \rangle=0$. Hence since $[\rho',H]=0 \implies [\rho',H_i]=0$ (because $H$ and $H_i$ have a common eigenbasis) we have $[\rho',H]=0 \implies \delta \langle \Delta E_i \rangle=0$. Hence when the post measurement state $\rho'$ commutes with $H$, all quantities $\langle \Delta E_i \rangle$, $\langle \Delta E \rangle$ are extremal, and so are the efficiencies $\eta^{[E]}= \langle \Delta E \rangle/\langle \Delta E_1 \rangle$, $\eta^{[R]}=-\langle \Delta E_2 \rangle/\langle \Delta E \rangle$.

\section{Operation ranges\label{app:ranges}}

As seen in Sec. \ref{app:optimal}, in the range $\frac{\beta_1}{\beta_2}\leq \frac{\omega_2}{\omega_1}\leq 1$, the minimal value of $\langle\Delta E\rangle$ over all bi-stochastic matrces $P$ is non-positive. Similarly, namely by the Birkhoff/von Neumann expansion and direct evaluation of the expansion coefficients, one obtains as well that the minimal value of $ \langle\Delta E_1\rangle$ is non-positive and the minimal value of $\langle\Delta E_2\rangle$ is null.
So in this region refrigeration (for which $\langle\Delta E_2\rangle\leq 0$) is forbidden  and the possible operations are energy extraction, thermal accelerator and heater.\\

Likewise, as seen in Sec. \ref{app:optimal}, in the range $\frac{\omega_2}{\omega_1}\leq 1\leq \frac{\beta_1}{\beta_2}\leq 1$, the minimal value of $\langle\Delta E_2\rangle$ is non-positive. Similarly one can also see that the minimal value of $ \langle\Delta E\rangle$ and of $\langle\Delta E_1\rangle$ are both null.
So in this region energy extraction (for which $\langle\Delta E\rangle\leq0$) and thermal accelerator (for which $\langle\Delta E_1\rangle\leq0$) are forbidden and the possible operations are refrigerator and heater.\\

Finally, in the region $\frac{\omega_2}{\omega_1}\geq 1$, one can see, by means of similar reasoning, that 
the minimal values of $\langle\Delta E\rangle$ and$ \langle\Delta E_2\rangle$ are null, and the minimal value of $\langle\Delta E_1\rangle$ is non-positive. Accordingly, energy extraction and refrigeration are forbidden ($\langle\Delta E\rangle$ and $\langle\Delta E_1\rangle$ can't be negative) and the possible operations are thermal accelerator and heater.

We remark that the minimal values of $\langle \Delta E\rangle, \langle\Delta E_1\rangle$ and $\langle\Delta E_2\rangle$ over all possible doubly stochastic matrices $P$, coincide with the negative ergotropies  \cite{Allahverdyan04EPL67} $\mathcal W_H, \mathcal W_{H_1}, \mathcal W_{H_2}$, relative to the Hamiltonians $H$, $H_1$, $H_2$, where the ergotropy $\mathcal W_K$, relative to some Hamiltonian $K$ reads
\begin{align}
\mathcal{W}_K = \text{min}_U \Tr (U\rho U^\dagger- \rho)K\, .
\end{align}
At the points $\omega_2/\omega_1=\beta_1/\beta_2$ and $\omega_2/\omega_1=1$ an exchange of ordering of the eigenvalues of the initial state $\rho$, occurs, which may accompany a passage of the state from passive to active, with respect to one or more of the Hamiltonians $H$, $H_1$, $H_2$. As a result 
the $H$-ergotropy $\mathcal{W}_H$ is null outside the range  $\frac{\beta_1}{\beta_2}\leq \frac{\omega_2}{\omega_1}\leq 1$;
the $H_1$-ergotropy $\mathcal{W}_{H_1}$ is null outside the region $\frac{\omega_2}{\omega_1}\geq \frac{\beta_1}{\beta_2}$; while
 the $H_2$-ergotropy $\mathcal{W}_{H_2}$ is null outside the region $\frac{\omega_2}{\omega_1}\leq \frac{\beta_1}{\beta_2}$.

\section{Group average of energy expectation changes\label{app:group average}}
In the following we shall calculate the quantities $\overline{\langle \Delta E_i \rangle}$ where the overline denotes average over the invariant measure of $SU(4)$. We use the parametrisation of the group in terms of generalised Euler angles $\boldsymbol{\alpha}=(\alpha_1, \alpha_2, \dots ,  \alpha_{15})$ following Ref. \cite{Tilma02JPA48}. The general element $U_{\boldsymbol{\alpha}}$ of $SU(4)$ reads \cite{Tilma02JPA48}:
\begin{align}
U_{\boldsymbol{\alpha}} = e^{i\lambda_3 \alpha_1}e^{i\lambda_2 \alpha_2}e^{i\lambda_3 \alpha_3}e^{i\lambda_5 \alpha_4}e^{i\lambda_3 \alpha_5}e^{i\lambda_{10} \alpha_6}e^{i\lambda_3 \alpha_7}e^{i\lambda_2 \alpha_8}
e^{i\lambda_3 \alpha_{9}}e^{i\lambda_5 \alpha_{10}}e^{i\lambda_3 \alpha_{11}}e^{i\lambda_2 \alpha_{12}}e^{i\lambda_3 \alpha_{13}}e^{i\lambda_8 \alpha_{14}}e^{i\lambda_{15} \alpha_{15}}
\end{align}
where $\lambda_i$ are the Gell-Mann matrices \cite{Tilma02JPA48}
\begin{equation}
\begin{array}{crcr}
\lambda_1 = \left( \begin{array}{cccc}
                     0 & 1 & 0 & 0 \\
                     1 & 0 & 0 & 0 \\
                     0 & 0 & 0 & 0 \\
                     0 & 0 & 0 & 0  \end{array} \right), &
\lambda_2 = \left( \begin{array}{crcr} 
                     0 & -i & 0 & 0 \\
                     i &  0 & 0 & 0 \\
                     0 &  0 & 0 & 0 \\
                     0 &  0 & 0 & 0  \end{array} \right), &
\lambda_3 = \left( \begin{array}{crcr} 
                     1 &  0 & 0 & 0 \\
                     0 & -1 & 0 & 0 \\
                     0 &  0 & 0 & 0 \\
                     0 &  0 & 0 & 0  \end{array} \right), \\
\lambda_4 = \left( \begin{array}{clcr} 
                     0 & 0 & 1 & 0 \\
                     0 & 0 & 0 & 0 \\
                     1 & 0 & 0 & 0 \\
                     0 & 0 & 0 & 0  \end{array} \right), &
\lambda_5 = \left( \begin{array}{crcr} 
                     0 & 0 & -i & 0 \\
                     0 & 0 &  0 & 0 \\
                     i & 0 &  0 & 0 \\
                     0 & 0 &  0 & 0 \end{array} \right), &
\lambda_6 = \left( \begin{array}{crcr} 
                     0 & 0 & 0 & 0 \\
                     0 & 0 & 1 & 0 \\
                     0 & 1 & 0 & 0 \\
                     0 & 0 & 0 & 0 \end{array} \right), \\
\lambda_7 = \left( \begin{array}{crcr} 
                     0 & 0 &  0 & 0 \\
                     0 & 0 & -i & 0 \\
                     0 & i &  0 & 0 \\
                     0 & 0 &  0 & 0 \end{array} \right), &
\lambda_8 = \frac{1}{\sqrt{3}}\left( \begin{array}{crcr} 
                     1 & 0 &  0 & 0 \\
                     0 & 1 &  0 & 0 \\
                     0 & 0 & -2 & 0 \\
                     0 & 0 &  0 & 0  \end{array} \right), &
\lambda_9 = \left( \begin{array}{crcr} 
                     0 & 0 & 0 & 1 \\
                     0 & 0 & 0 & 0 \\
                     0 & 0 & 0 & 0 \\
                     1 & 0 & 0 & 0 \end{array} \right), \\
\lambda_{10} = \left( \begin{array}{crcr} 
                     0 & 0 & 0 & -i \\
                     0 & 0 & 0 &  0 \\
                     0 & 0 & 0 &  0 \\
                     i & 0 & 0 &  0  \end{array} \right), &
\lambda_{11} = \left( \begin{array}{crcr} 
                     0 & 0 & 0 & 0 \\
                     0 & 0 & 0 & 1 \\
                     0 & 0 & 0 & 0 \\
                     0 & 1 & 0 & 0\end{array} \right), &
\lambda_{12} = \left( \begin{array}{crcr} 
                     0 & 0 & 0 &  0 \\
                     0 & 0 & 0 & -i \\
                     0 & 0 & 0 &  0 \\
                     0 & i & 0 &  0  \end{array} \right), \\
\lambda_{13} = \left( \begin{array}{crcr} 
                     0 & 0 & 0 & 0 \\
                     0 & 0 & 0 & 0 \\
                     0 & 0 & 0 & 1 \\
                     0 & 0 & 1 & 0  \end{array} \right), &
\lambda_{14} = \left( \begin{array}{crcr} 
                     0 & 0 & 0 &  0 \\
                     0 & 0 & 0 &  0 \\
                     0 & 0 & 0 & -i \\
                     0 & 0 & i &  0  \end{array} \right), &
\lambda_{15} = \frac{1}{\sqrt{6}}\left( \begin{array}{crcr}
                     1 & 0 & 0 &  0 \\
                     0 & 1 & 0 &  0 \\
                     0 & 0 & 1 &  0 \\
                     0 & 0 & 0 & -3  \end{array} \right).
\end{array}    
\end{equation}

The group average of a function $g(\boldsymbol{\alpha})$ reads:
\begin{align}
\overline{g} = \int d\mathcal{M}(\boldsymbol{\alpha}) g(\boldsymbol{\alpha})\, ,
\end{align}
where 
\begin{align}
\label{dvsu4}
d\mathcal{M}(\boldsymbol{\alpha}) = \frac{\cos(\alpha_{4})^3\cos(\alpha_{6})\cos(\alpha_{10})\sin(2\alpha_{2})\sin(\alpha_{4})
\sin(\alpha_{6})^5\sin(2\alpha_{8})\sin(\alpha_{10})^3\sin(2\alpha_{12})d\alpha_{15}\ldots d\alpha_{1}}
{\int \cos(\alpha_{4})^3\cos(\alpha_{6})\cos(\alpha_{10})\sin(2\alpha_{2})\sin(\alpha_{4})
\sin(\alpha_{6})^5\sin(2\alpha_{8})\sin(\alpha_{10})^3\sin(2\alpha_{12})d\alpha_{15}\ldots d\alpha_{1}}
\end{align}
is the normalised Haar measure of the group, and the integration is carried over the ranges:
\begin{gather}
0 \le \alpha_1,\alpha_7,\alpha_{11} \le \pi;
0 \le \alpha_2,\alpha_4,\alpha_6,\alpha_8,\alpha_{10},\alpha_{12}
\le \frac{\pi}{2};
0 \le \alpha_3,\alpha_5,\alpha_9,\alpha_{13} \le 2\pi;
0 \le \alpha_{14} \le \sqrt{3}\pi;0 \le \alpha_{15} \le 2\sqrt{\frac{2}{3}}\pi.
\end{gather}

The energy changes associated to a random basis  $|\psi_k(\boldsymbol{\alpha})\rangle$, obtained by applying a random unitary $U_{\boldsymbol{\alpha}}$ to the energy eigenbasis $|\psi_k(\boldsymbol{\alpha})\rangle=U_{\boldsymbol{\alpha}} |k\rangle$, read:
\begin{align}
\langle \Delta E_i \rangle(\boldsymbol{\alpha})&= \int d\Delta E_1d\Delta E_2 P(\Delta E_1,\Delta E_2) \Delta E_i = \sum_{mn} E^i_m (q_{mn}(\boldsymbol{\alpha})-\delta_{mn}) p_n^0 =
 \Tr H_i (\Phi_{\boldsymbol{\alpha}}[\rho]-\rho)\, ,
\end{align}
where
\begin{align}
q_{m|n}(\boldsymbol{\alpha}) & = \sum_k \Tr\,  \Pi_m \pi_k(\boldsymbol{\alpha}) \Pi_n \pi_k(\boldsymbol{\alpha}) \Pi_m, \qquad \pi_k(\boldsymbol{\alpha}) = U_{\boldsymbol{\alpha}} \pi_k U_{\boldsymbol{\alpha}}^\dagger, \qquad
\Phi_{\boldsymbol{\alpha}}[\rho]= \sum_k U_{\boldsymbol{\alpha}}  \Pi_k U^\dagger_{\boldsymbol{\alpha}}  \rho U_{\boldsymbol{\alpha}}  \Pi_k U^\dagger_{\boldsymbol{\alpha}}\, .
\end{align}
Hence:
\begin{align}
\overline{\langle \Delta E_i \rangle}&= \Tr H_i (\overline{\Phi}[\rho]-\rho)\, ,
\\
\overline{\Phi}[\rho]&= \int d\mathcal{M}({\boldsymbol{\alpha}} )\sum_k U_{\boldsymbol{\alpha}}  \Pi_k U^\dagger_{\boldsymbol{\alpha}}  \rho U_{\boldsymbol{\alpha}}  \Pi_k U^\dagger_{\boldsymbol{\alpha}}\, .
\end{align}

Together with the $4\times 4$ identity matrix, the $15$ traceless Gell-Mann matrices form a basis for Hermitian $4\times 4$ matrices. Therefore any $4\times 4$ density matrix $\rho$ can be expressed as:
\begin{align}
\rho= \mathbb{1}/4 + \mathbf{n} \cdot \boldsymbol{\lambda} 
\end{align}
where $\boldsymbol{\lambda} = (\lambda_1, \lambda_2 , \dots , \lambda_{15})^T$ and $\mathbf{n} \in \mathbb{R}^{15}$, with $|\mathbf{n}|\leq \sqrt{3/8}$ (ensuring $\Tr \rho^2 \leq 1$). This expression generalises the Bloch representation of $1/2$-spin. 

We have
\begin{align}
\overline{\Phi}[\rho] &= \int d\mathcal{M}(\boldsymbol{\alpha})\sum_k U_{\boldsymbol{\alpha}}  \Pi_k U^\dagger_{\boldsymbol{\alpha}}  \rho U_{\boldsymbol{\alpha}}  \Pi_k U^\dagger_{\boldsymbol{\alpha}} = \mathbb{1}/4 +  \int d\mathcal{M}(\boldsymbol{\alpha})\sum_k U_{\boldsymbol{\alpha}}  \Pi_k U^\dagger_{\boldsymbol{\alpha}}  
\mathbf{n} \cdot \boldsymbol{\lambda}
 U_{\boldsymbol{\alpha}}  \Pi_k U^\dagger_{\boldsymbol{\alpha}}\, .
\end{align}
The effect of $U_{\boldsymbol{\alpha}}$ on a state is to ``rotate'' the ``Bloch'' vector $\mathbf{n}$, let us denote that rotation with $R_{\boldsymbol{\alpha}}$:
\begin{align}
U^\dagger_{\boldsymbol{\alpha}} \mathbf{n} \cdot \boldsymbol{\lambda} U_{\boldsymbol{\alpha}}  = R_{\boldsymbol{\alpha}}\mathbf{n}  \cdot \boldsymbol{\lambda} \, .
\end{align}
Now let us focus on the effect of the projectors $\Pi_k$ on $\mathbf{n}\cdot \boldsymbol{\lambda}$, that is let us consider the expression $
\sum_k \Pi_k ( \mathbf{n} \cdot \boldsymbol{\lambda}) \Pi_k
$.
The projectors $\Pi_k$ are of the form $diag(0,0,1,0)$ with $1$ at the $k$-th position, therefore their effect on a matrix is to erase the off diagonal term. By inspection of the explicit form of the Gell-Mann matrices $\lambda_i$, we have
\begin{align}
\left\{\begin{array}{ll}
\sum_k \Pi_k \lambda_i \Pi_k= \lambda_i & \text{ for } i=3,8,15  \\
\sum_k \Pi_k \lambda_i \Pi_k = 0  & \text{ otherwise}.
\end{array}\right.
\end{align}
Therefore
\begin{align}
\sum_k \Pi_k ( \mathbf{n} \cdot \boldsymbol{\lambda}) \Pi_k = Q\mathbf{n} \cdot \boldsymbol{\lambda}\, ,
\end{align}
where $Q \mathbf{n} = (0,0,n_3,0,0,0,0,n_8,0,0,0,0,0,0,n_{15})$ is the projection of $\mathbf{n}$ onto the subspace spanned by $n_3,n_8,n_{15}$. 
Combining everything:
\begin{align}
\overline{\Phi}[\rho] = \mathbb{1}/4 + \left( \int d\mathcal{M}(\boldsymbol{\alpha})R_{\boldsymbol{\alpha}}^T  Q R_{\boldsymbol{\alpha}} \mathbf{n}   \right ) \cdot \boldsymbol{\lambda}= \mathbb{1}/4 + A \mathbf{n} \cdot \boldsymbol{\lambda}\, ,
\end{align}
where
\begin{align}
A = \int d\mathcal{M}(\boldsymbol{\alpha})R_{\boldsymbol{\alpha}}^T  Q R_{\boldsymbol{\alpha}}\, ,
\end{align}
and the integrand $R_{\boldsymbol{\alpha}}^T  Q R_{\boldsymbol{\alpha}} $ represents the projector onto the rotated subspace spanned by $n_3,n_8,n_{15}$.
Let us consider the projector $Q_i$ onto the space spanned by $n_i$ and let 
\begin{align}
M_i=\int d\mathcal{M}(\boldsymbol{\alpha})R_{\boldsymbol{\alpha}}^T  Q_i R_{\boldsymbol{\alpha}}\, .
\end{align}
We note that $Q=Q_3+Q_8+Q_{15}$, hence $A= M_3+M_8+M_{15}$.
By spherical symmetry, $M_i$ does not depend on $i$, $M_i=M$, and, by completeness (i.e., $\sum_i Q_i=\mathbb{I}$),
 $\sum_i M_i = \mathbb{I}=15 M $ (here $\mathbb{I}$ is the identity in $\mathbb{R}^{15}$ not to be confused with the symbol $\mathbb{1}$ above), hence $M = \mathbb{I}/15$,  $A= \mathbb{I}/5$. Thus:
\begin{align}
\overline{\Phi}[\rho] = \frac{\mathbb{1}}{4} +  \frac{\mathbf{n}}{5} \cdot \boldsymbol{\lambda}\, .
\end{align}
In general for $SU(N)$ one would obtain $\overline{\Phi}[\rho] = \mathbb{1}/N +  \mathbf{n} \cdot \boldsymbol{\lambda}/(N+1)$; because $1/(N+1)$ is the ratio of the number of diagonal Gell-Mann matrices $\boldsymbol{\lambda}$, i.e. $N-1$, over the total number of Gell-Matrices, i.e. $N^2-1$.

Let us expand the (traceless) Hamiltonian of qubit $i$, $H_i$ over the Gell-Mann basis:
\begin{align}
H_i = \mathbf{h}_i\cdot  \boldsymbol{\lambda}\, ,
\end{align}
where $\mathbf{h}_i \in \mathbb{C}^{15}$.
Then, the change in energy expectation of qubit $i$ reads:
\begin{align}
\overline{\langle \Delta E_i \rangle} &= \Tr  (\mathbb{1}/4 +\mathbf{n}/5 \cdot  \boldsymbol{\lambda} - \mathbb{1}/4 - \mathbf{n} \cdot  \boldsymbol{\lambda})(\mathbf{h}_i\cdot  \boldsymbol{\lambda})
= -(4/5)\Tr\, (\mathbf{n} \cdot  \boldsymbol{\lambda})( \mathbf{h}_i\cdot  \boldsymbol{\lambda})
= -(4/5)\Tr\, (\mathbb{1}/4+ \mathbf{n} \cdot  \boldsymbol{\lambda})( \mathbf{h}_i\cdot  \boldsymbol{\lambda}) \nonumber\\
&=-(4/5) \langle E_i \rangle_{in}\, .
\end{align}
Since the initial state of subsystem $i$ is at positive temperature, i.e. $\beta_i > 0$ it is $\langle E_i \rangle_{in}  \leq0$, hence 
 $\overline{\langle \Delta E_i \rangle}\geq 0$.

\begin{figure}[t] 
\includegraphics[width=250pt]{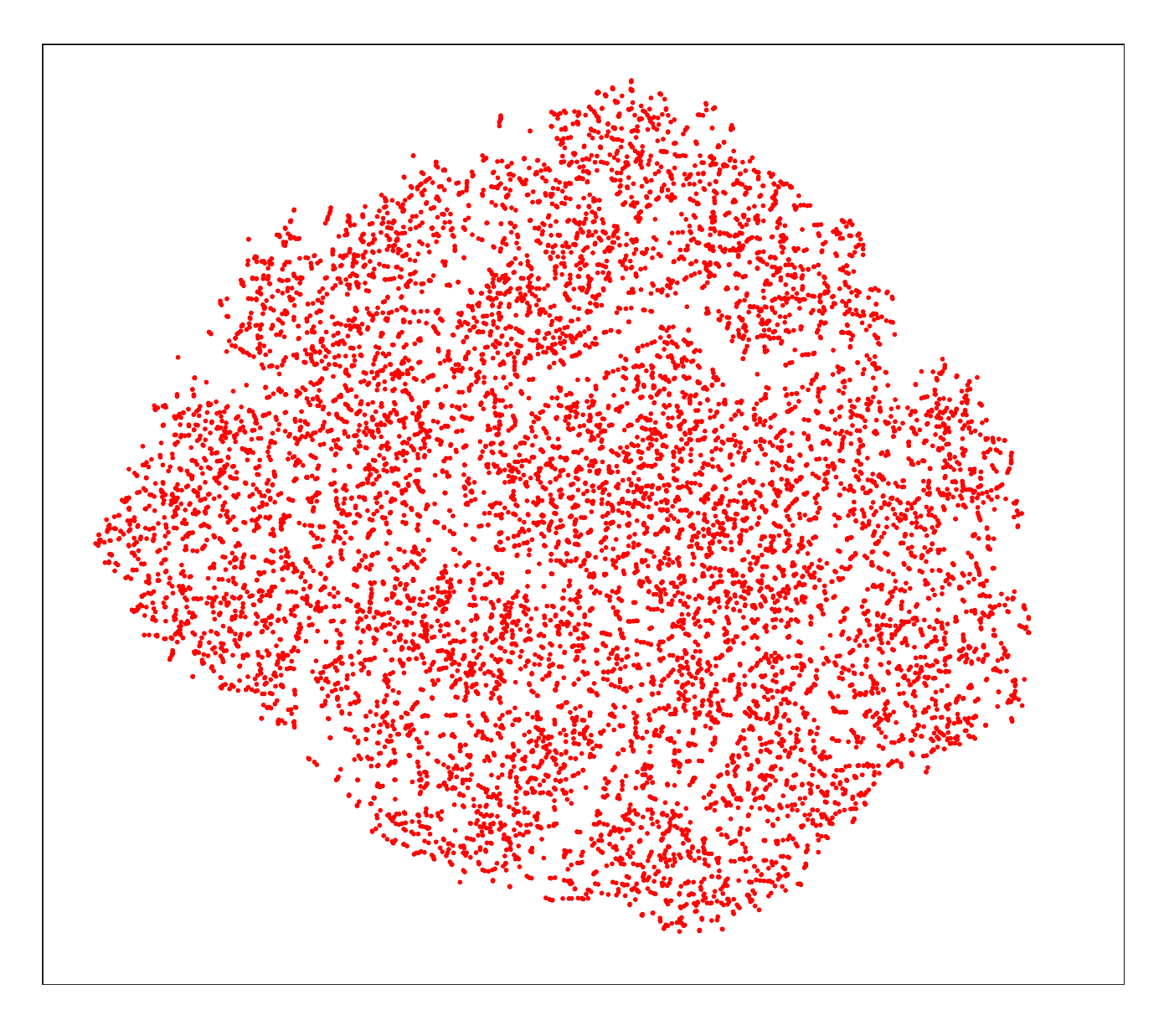}
\caption{2-dimensional representation of $\mathcal{R}_{[R]}$ obtained by the t-SNE algorithm \cite{tsne}. The points in the image are a representative sample of $10^4$ elements in the $\mathcal{R}_{[R]}$ region, the data was taken from a Monte Carlo simulation performed at $\beta_1/\beta_2=2/3$ and $\omega_2/\omega_1=0.6$. 
}
\label{fig:connect}
\end{figure}

\section{Connectedness of the subset of $SU(4)$ for which QMC occurs
\label{app:connect}}
Let $\mathcal{R}_{[R]}= \{ U \in SU(4) | \Tr H_2(\sum U\Pi_kU^\dagger \rho U\Pi_kU^\dagger-\rho)\leq 0\}$ be the subset of $SU(4)$ in which quantum measurement cooling is realized. We have performed a numerical analysis aimed at assessing whether $\mathcal{R}_{[R]}$ is connected. To this end we have used a dimensionality reduction algorithmm that is common in machine learning, namely the t-distributed Stochastic Neighbors Embedding (t-SNE)  \cite{tsne}. This algorithm maps high dimensional data onto a lower dimensional space (of 2 or 3 dimensions) preserving the local spatial structure of the data. Namely, if the high-dimensional data are clustered they will be clustered in the low dimensional representation as well. Fig. \ref{fig:connect} shows the mapping onto a 2 dimensional space of the subset of unitaries belonging to $\mathcal{R}_{[R]}$ from a sample of elements of $SU(4)$ uniformly sampled from the Eucledian measure $d\boldsymbol{\alpha}$. The data does not appear clustered, thus providing strong evidence that $\mathcal{R}_{[R]}$ is connected. The same consideration applies for the regions of the $SU(4)$ space belonging to the other regimes. Details about the MATLAB implementation of the algorithm, used to produce Fig. \ref{fig:connect} can be found in \cite{tsne}.

\end{document}